\documentclass[referee]{raa}            

\usepackage{graphicx,times}             
\usepackage{natbib}
\usepackage{amssymb,amsmath}
\usepackage[T1]{fontenc}
\usepackage{url}
\bibpunct{(}{)}{;}{a}{}{,}

\begin{document}

   \title{Photometric and Spectroscopic study of Ten Low Mass Ratio Contact Binary Systems: Orbital Stability, O'Connell Effect and Infra-red Calcium Line Filling.
}

   \volnopage{Vol.0 (20xx) No.0, 000--000}      
   \setcounter{page}{1}          

   \author{Surjit S. Wadhwa
      \inst{1}
      \and Adam Popowicz
        \inst{2}
        \and  Raúl Michel
         \inst{3}
         \and Petar Kostić
         \inst{4}
         \and Oliver Vince
         \inst{4}
       \and Nick F. H. Tothill
        \inst{1}
   \and Ain Y. De Horta
      \inst{1}
   \and Miroslav D. Filipovi\'c
      \inst{1}
}

   \institute{School of Science, Western Sydney University, Locked Bag 1797, Penrith, NSW 2751, Australia; {\it 19899347@student.westernsydney.edu.au}\\
        \and
        Department of Electronics, Electrical Engineering and Microelectronics, Silesian University of Technology, Akademicka 16, 44120 Gliwice, Poland
        \and
        Instituto de Astronom\'{\i}a, UNAM. A.P. 877, 22800 Ensenada, BC, M\'exico
        \and
             Astronomical Observatory, Volgina 7, 11060 Belgrade, Serbia\\
\vs\no
   {\small Received~~20xx month day; accepted~~20xx~~month day}}

\abstract{ Low mass ratio contact binary systems are more likely to have unstable orbits and potentially merge. In addition, such systems exhibit characteristics such as starspots and high energy emissions (UV) suggestive of chromospheric and magnetic activity. Light curve modelling of ten contact binary systems is reported. All were found to be of extreme low mass ratio ranging from 0.122 to 0.24 and three were found to be potentially unstable and possible merger candidates. Filling of the infrared Calcium absorption lines is a marker of increased chromospheric activity. We use the available LAMOST spectra along with matched standard spectra (broadened for rotation) to measure the excess filling of the central core depression flux of the two main infrared Calcium absorption lines $\lambda 8542$ and $\lambda 8662$. We find that all reported contact binaries have excess filling of the core flux in the infrared Calcium lines. Three of the systems reported were also observed by the GALEX mission and we find that all three have features of excess ultraviolet emissions further adding evidence for increased chromospheric activity in low mass ratio contact binaries. Analysis of both orbital stability and absorption line filling is dependent on the determination of geometric and absolute parameters from light curve modelling. Not an insignificant number of contact binary light curves exhibit the O’Connell effect, usually attributed to starspots. We discuss the inclusion of starspots in light curve solutions and how they influence the geometric and absolute parameters. 
}

   \authorrunning{Wadhwa et al}            
   \titlerunning{Infra-red CaII filling in contact binaries}  

   \maketitle

%
%
\section{Introduction}
 \label{sec:intro}

Contact binaries usually comprise of a low mass (spectral class F to K) primary with an even lower mass secondary. Both stars are distorted such that they fill their respective Roche lobes and are surrounded by a common envelope. The primary acts much like a main sequence star of similar mass and there is significant transfer of energy between the components such that the common envelope usually has temperature similar to that of the primary component \citep{2013MNRAS.430.2029Y}. The orbital period ranges from about 0.2 to 1 day with most systems having periods between 0.3 and 0.4 days \citep{2021ApJS..254...10L}. The orbit is thought to be circular and synchronous. High orbital and rotation speeds combined with the common envelope geometry usually results in significantly increased magnetic and chromospheric activity manifesting as high energy (UV) excess and potentially starspots, flares, accretion gas impacts or Coriolis heating \citep{2017MNRAS.465.4678M}. In addition, such increased magnetic activity is postulated to result in magnetic breaking and orbital instability leading to potential merger of the components. Theoretical considerations suggests that for low mass contact binaries, merger is most likely at very low mass ratios \citep{2007MNRAS.377.1635A, 2009MNRAS.394..501A}.

Most contact binary systems show some type of choromospheric and coronal activity signals. The most common photometric phenomenon is the asymmetry in the light curve maxima (O'Connell effect) thought to be the result of starspots. Significant emission excesses have also been noted particularly at shorter wavelengths (X-Ray, UV, MgII, H$\alpha$, CaII) \citep{2020A&A...635A..89M}. Unfortunately, optical spectral evaluation of the Magnesium, Hydrogen and Calcium emissions requires high resolution imaging due to the presence of other nearby emission lines. The infrared part of the spectrum is relatively quiet and offers potential to review chromospheric spectral changes in contact binaries with low to medium resolution spectra. \citet{2022AJ....164..202L, 2023MNRAS.519.5760L} have published results of potential filling of the infrared H$\alpha$ absorption line as a marker of increased chromospheric activity. In this study we explore the utility of filling of the infrared CaII $\lambda$8542 and $\lambda$8662 lines from the The Large Sky Area Multi-Object Fibre Spectroscopic Telescope (LAMOST) survey spectra as a potential marker of increased chromospheric activity in low mass ratio contact binary systems. In addition, we briefly look at increased ultraviolet emissions as a marker for increased chromospheric activity in three systems.

Ever since the confirmation that luminous red novae are likely the product of contact binary merger \citep{2011A&A...528A.114T} the topic of orbital stability of such systems has been an active area of investigation. Theoretical considerations have demonstrated that merger is likely to occur at very low mass ratios however there is no global minimum mass ratio at which merger will occur with each system having a unique instability value depending on its Roche geometry \citep{2021MNRAS.501..229W}. Application of these theoretical considerations has resulted in the identification of over 20 potential merger candidates \citep{2021MNRAS.501..229W, 2022RAA....22j5009W, 2023PASP..135g4202W}. We use recently published \citep{2022JApA...43...94W} techniques to identify ten likely extreme low mass ratio potential merger candidates. In addition, we use the LAMOST survey medium resolution spectra to characterise the filling of infrared CaII absorption lines for nine of these systems. Since the initial start of the current study, partial analysis of two systems reported in this study have been published and we discuss our review considering the published findings.

\section{Observations}
 \label{sec:S2}
 \subsection{Image acquisition and light curves}
 The basic identification details, abbreviations (used going forward) along with comparison and check stars are summarised in Table 1. A0304, A0727, A0748, A0806 and IT Cnc were observed with the Las Cumbres Observatory (LCO) network of 40cm telescopes equipped with either the SBIG STL6303 camera with resolution of 0.571 arcsec/pixel or QHY600 CMOS camera with resolution of 0.73 arcsec/pixel. Both cameras are equipped with standard photometric filters. 
 A0346 was observed at the  Vidojevica observatory with the 60cm Telescope Nedeljkovi\'c equipped with FLI ProLine PL230 CCD camera with a resolution of 0.52 arcsec/pixel and standard filters. Some imaging of A0346 was also performed with the LCO network. V1359 Cas, N6978 and N3650 were observed with Silesian University of Technology Observatories {(SUTO\footnote{\url{www.suto.aei.polsl.pl}})} 0.3m remotely operated telescope equipped with ASI ZWO1600MM cooled CMOS camera with resolution of 0.7 arcsec/pixel. PS Com was observed at the San Pedro M\'artir Observatory using the 84cm Ritchey-Chretien telescope equipped with Marconi 5 CCD camera with resolution of 0.25 arcsec/pixel. All systems were observed using standard V band filters (Bessel or Johnston).

 \begin{table*}
    \centering
    \caption{Name, Abbreviation (used in text), comparison star and check star identifiers for the ten contact binary systems.}
    \begin{tabular}{|c|c|c|c|}
    \hline
        Name & Abbrv & Comparison Star & Check Star \\ \hline
        NSVS 3650324 & N3650 & TYC 2790-1689-1 & 2MASS J00141483+4151308 \\ 
        V1359 Cas & V1359 Cas & 2MASS J02525135+6628000 & 2MASS J02530271+6631067 \\
        ASAS J030424+0611.8 & A0304 & 2MASS J03034648+0620469 & 2MASS J03040675+0608231 \\ 
        ASASSN-V J034633.63+410815.8 & A0346 & 2MASS J03461525+4107406 & 2MASS J03461737+4109029 \\ 
        NSVS 6798913 & N6798 & HD279807 & 2MASS J04255488+3658220 \\ 
        ASAS J072718+0837.8 & A0727 & UCAC4 494-043214 & 2MASS J07271257+0835408 \\ 
        ASAS J074829+1904.1 & A0748 & TYC 1370-475-1 & 2MASS J07482414+1859391 \\ 
        ASAS J080638+1150.4 & A0806 & 2MASS J08063654+1154126 & 2MASS J08064318+1147557 \\ 
        IT Cnc & IT Cnc & TYC 1359-656-1 & 2MASS J08424202+2126055 \\ 
        PS Com & PS Com & 2MASS J11585295+1414392 & 2MASS J11591312+1410221 \\ \hline
    \end{tabular}
\end{table*}
 
  Aperture photometry was performed using the AstroImage J software package \citep{2017AJ....153...77C} with comparison and check stars as indicated in Table 1. Comparison and check star magnitudes were adopted from the AAVSO Photometric All-Sky Survey \citep{2015AAS...22533616H}. The software package provides an estimate of the photometric error and all observations with estimated error greater than 0.01 magnitude were excluded. Photometric data was folded using published ephemera from the All Sky Automated Survey - Super Novae (ASAS-SN) \citep{2014ApJ...788...48S, 2018MNRAS.477.3145J}. Basic light curve characteristics are summarised in Table 2. Only two systems, A0727 and PS Com, demonstrated the O'Connell effect of greater than 0.01 magnitude. Given the lack of high cadence historical observations a formal period study could not be accurately performed on the systems.

  \begin{table}
    \centering
    \caption{Light curve V band magnitudes of the ten systems observed. V1(Max) = first maximum after the primary eclipse (= phase 0.25), V2(max) = second maximum after the secondary eclipse (= phase 0.75), V(Sec) = secondary eclipse and V(Primary) = Primary Eclipse. Epoch (+2450000) and Period adopted from ASAS-SN}
    \begin{tabular}{|c|c|c|c|c|c|c|}
    \hline
        SYSTEM & V1 (Max) & V2 (Max) & V (Sec) & V (Primary)&Epoch (HJD)&Period (d) \\ \hline
        N3650 & 11.82 & 11.81 & 12.15 & 12.21 &6454.06703&0.3792974\\ 
        V1359 Cas & 12.12 & 12.13 & 12.37 & 12.39&8430.93255&0.3602919 \\ 
        A0304 & 11.59 & 11.59 & 11.94 & 11.97&7385.78282&0.2753578 \\ 
        A0346 & 13.37 & 13.37 & 13.70 & 13.72&7422.79980&0.3071700 \\ 
        N6798 & 11.37 & 11.37 & 11.67 & 11.73&7766.75601&0.3625408 \\ 
        A0727 & 12.12 & 12.06 & 12.47 & 12.48&7397.84334&0.315709 \\ 
        A0748 & 12.67 & 12.66 & 13.03 & 13.03&7802.88997&0.3055999 \\ 
        A0806 & 12.04 & 12.04 & 12.36 & 12.41&8046.85314&0.2974811 \\ 
        IT Cnc & 12.43 & 12.44 & 12.73 & 12.74&7036.07919&0.3636616 \\ 
        PS Com & 12.88 & 12.92 & 13.34 & 13.40&8117.08614&0.2527505 \\ \hline
    \end{tabular}
\end{table}
  
  \subsection{Mass of the primary component}
  Apart from high resolution spectral radial velocity observations there is no direct way to measure the mass of the component stars in a binary system. As the primary component acts like a main sequence star \citep{2013MNRAS.430.2029Y}, investigators have used colour or other empirical relationships such as period-luminosity \citep{2021AJ....162...13L} or period-separation \citep{2022AJ....164..202L} to estimate the mass of the primary ($M_1$). Release of highly accurate distance estimates from the GAIA mission \citep{2022A&A...658A..91A} affords an opportunity to estimate absolute magnitude and mass more directly from observations. We estimate the mass of the primary as the mean of an infrared colour calibration, and one based on the observed absolute magnitude of the primary ($M_{V1}$). As all systems have very low mass ratios (see below) the apparent magnitude at phase 0.5 (mid eclipse) represents the apparent magnitude of the primary component. As described in \citet{2023PASP..135g4202W} we adjust the apparent magnitude with a distance corrected extinction and then use the standard distance modulus and the GAIA distance to estimate the absolute magnitude of the primary component. We use the April 2022 update calibrations from \citet{2013ApJS..208....9P} for low mass main sequence stars to estimate the mass of the primary from the absolute magnitude estimate. The infrared colour calibration was chosen as measurements are little influenced by extinction and the observations by The Two Micron All SKY Survey (2MASS) \citet{2006AJ....131.1163S} were acquired simultaneously. We use the J-H colour from the 2MASS survey along with the same calibration tables to estimate the colour-based mass of the primary. We adopt the mean of the two estimates as the mass of the primary component. The photometric error has little influence on the estimation of absolute magnitude and does not introduce significant error to the estimation of the mass of the primary. As all systems are relatively close, the most distant A0806 has a distance estimate of less than 600pc, the GAIA error for the distances is quite low. Based on the GAIA distances the estimated mass of the primary varies by no more than 0.02$\rm M_{\sun}$ between the 16th and 85th percentiles of the distance estimate. We report this error in this study. There are likely to be errors associated with the calibration tables however these could not be reliably determined from the cited reference and we have not taken these into consideration. Where appropriate all other error estimates were propagated from this value. Primary component mass estimates for each system are summarised along with other absolute parameters (see below) in Table 4.

  Two systems, N6798 and N3650, have been observed previously \citep{2022AJ....164..202L}. The authors estimate the mass of the primary component using a period-separation relationship. Our estimates of the mass of the primary components are somewhat lower in both cases. We estimate the mass of the primary component of N6798 as $0.91M_{\sun}$ against their estimate of $1.48M_{\sun}$. We note the recently published StarHorse data for 8 spectroscopic surveys \citet{2023A&A...673A.155Q} estimates the mass of the system as between $0.96M_{\sun} - 1.02M_{\sun}$ while the GAIA DR 3 estimate is $1.01M_{\sun}$. We have chosen to adopt our estimate for the remainder of this study. In the case of N3650 the published estimate for the mass of the primary is $1.49M_{\sun}$ against our estimate of $1.18M_{\sun}$. As with N6798, StarHorse data for 8 spectroscopic surveys estimates the mass of N3650 as $1.18M_{\sun}$and the GAIA DR 3 estimated mass is $1.28M_{\sun}$ and we adopt our estimate for the remainder of this study.

  \section{Light Curve Analysis}
  We use the 2013 version of the Wilson-Devenney light curve modelling code (WD-Code) and the well described and accepted mass ratio grid search method to find the best fitting modelled light curve against the observed light curve \citep{1990ApJ...356..613W, 1998ApJ...508..308K, 2021NewA...8601565N}. The temperature of the primary component ($T_1$) is usually fixed during the grid search process. As all systems were observed with the LAMOST spectral survey we adopt the temperature reported from the seventh data release (DR7) \citep{2020RAA....20..163Q} except in the case of V1359 Cas where no DR7 data is available, we use published data from the eighth release (DR8) \citep{2023ApJS..266...40W}. As has been clearly shown, the absolute value of $\rm T_1$ has little influence on the geometric light curve solution and absolute parameter determinations are not greatly influenced if temperature-based calibrations are not adopted when determining them \citep{2023SerAJ.207...21W}. As all systems are of relatively low mass and have temperatures of less than $7000\rm K$, the bolometric albedoes were set equal $\rm A_{1,2} = 0.5$, gravity darkening coefficients were also equal $\rm g_{1,2} = 0.32$ and simple reflection treatment was applied. Limb darkening coefficients were interpolated from \citet{1993AJ....106.2096V}.

  During the grid search procedure, the temperature of the secondary component ($\rm T_2$), the fillout factor ($\rm f$), inclination ($\rm i$) and the scaling luminosity factor ($\rm L_1$) were the adjustable parameters. The initial mass ratio ($\rm q$) search was carried out for fixed values of $\rm q$ from 0.05 to 1.0 in steps of 0.05. The search grid was then refined in steps of 0.01 and 0.001 following the best fit from the previous search. In all cases iterations were carried out until the predicted correction was less than the reported standard deviation for all adjustable parameters. In the final iteration the mass ratio was also made an adjustable parameter and the reported standard deviations (one sigma) along with an adjustment for potential random error (see below) were adopted as the error for all parameters. In the cases of A0727 and PS Com, the two systems with significant O'Connell effect, we chose not to adopt a solution with starspots even though the fit was much better for reasons we explain below.  The light curve solutions for all systems are summarised in Table 3. Fitted and observed light curves are illustrated in Figure 1.

  \subsubsection{WD-Code and Error Estimation}
 Although it is well known that light curve solutions of contact binary systems with complete eclipses are highly accurate there has been some discussion with respect to the reported errors of the geometric solution. The Monte Carlo algorithm has been used in individual cases \citep{2024RAA....24e5001P} however there exists only one comprehensive error analysis study of contact binaries based on the WD-Code \citep{2021PASP..133h4202L}. The study by \citet{2021PASP..133h4202L} simulated the process of repeated measurements (Monte Carlo algorithm) for 48 different models of contact binaries taking into account variable photometric accuracy, variable timing accuracy, multiple values for the mass ratio, inclination, third light and temperatures of the secondary. The core aim of the study was to reduce the influence of random errors on the light curve solution. The main conclusions of the study were that in the absence of a third light and in a setting of complete eclipses the random errors in the solution are quite small. Higher cadence observations do not increase accuracy of the solution. The precision of the photometric measurement does have an influence mainly in cases where the eclipse is not total. In this study we err on the side of caution and report the one sigma value from the WD-Code plus the random error from the closest model (photometric precision of 0.01 magnitude, complete eclipses and no third light) from \citet{2021PASP..133h4202L}.

  \begin{figure}
    \label{fig:F1}
	\includegraphics[width=\columnwidth]{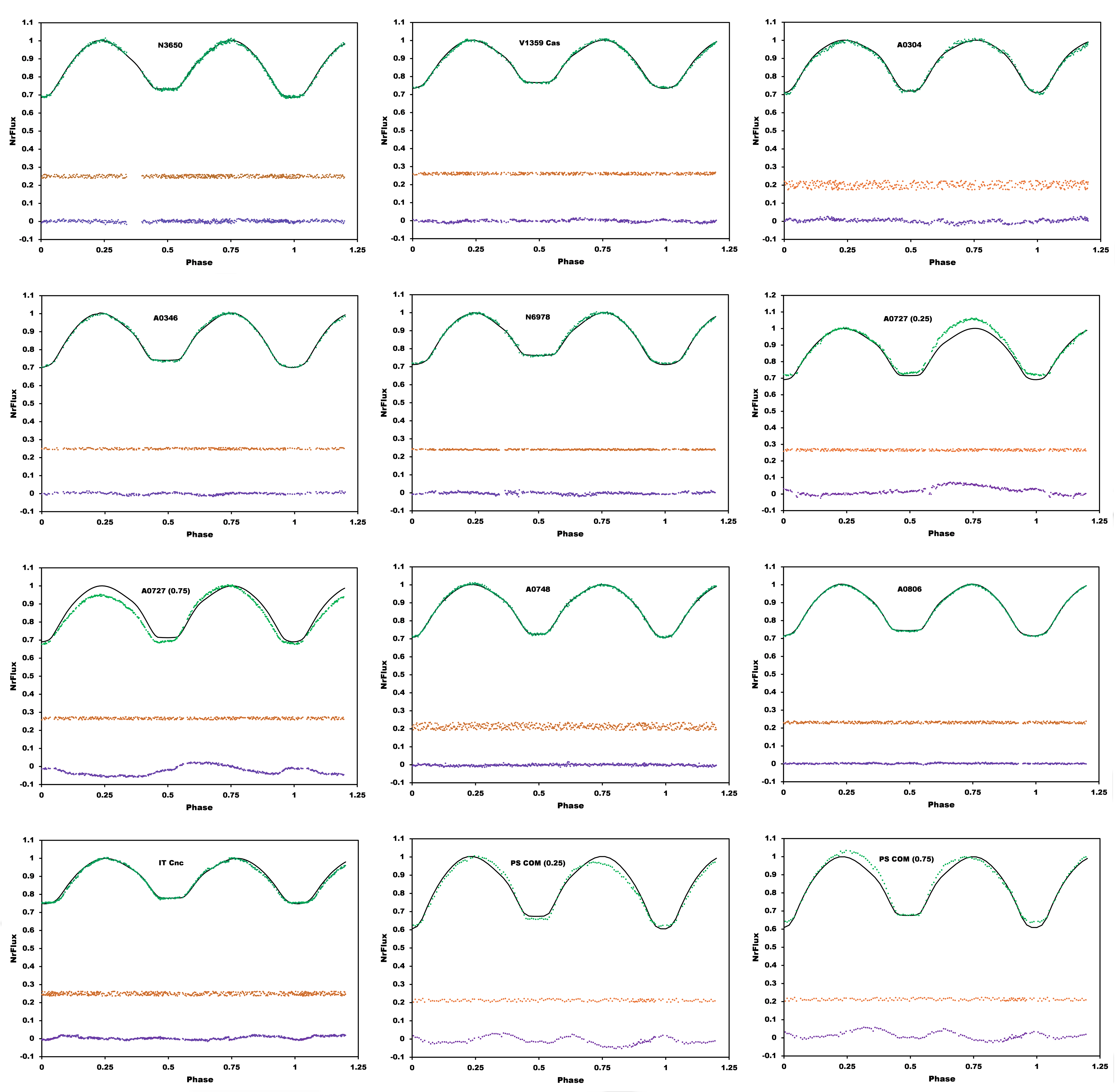}
    \caption{Observed and modelled light curves for the ten systems. The green light curves represent the V band and the black line represents the WD modelled curve. The orange curve represents the check star flux (vertically shifted for clarity) and the purple curve represents the residual between the observed (green) and modelled (black) curves. Both the phase 0.25 and 0.75 normalised curves are shown for the two systems exhibiting the O'Connell effect (A0727 and PS Com). The vertical axis labelled NrFlux represents the normalised flux.}
    \end{figure}

  \begin{table*}

   \centering
\caption{Light curve solution summary for the observed systems. ($q_{inst}$) Range = instability mass ratio range. For A0727 and PS Com solutions for maximum light at phases 0.25 and 0.75 are presented along with a mean solution.}
   \begin{tabular}{|c|c|c|c|c|c|c|}
    \hline
        \hfil Name &\hfil$T_1 (K)$ Fixed &\hfil $T_2 (K)$ &\hfil Incl $(^\circ)$&\hfil Mass Ratio ($q$) &\hfil Fillout (\%)   &\hfil $q_{inst}$ Range \\
        \hline
        \hfil N3560 &\hfil 6568 &\hfil $6411\pm21$  &\hfil $79.3\pm0.5$&\hfil $0.147\pm0.001$&\hfil $92\pm3$&\hfil 0.077 - 0.087 \\
        
        \hfil V1359 Cas   &\hfil 5944 &\hfil $5877\pm24$ &\hfil $79.7\pm0.6$&\hfil $0.123\pm0.001$&\hfil $54\pm5$&\hfil 0.121 - 0.147\\ 
         \hfil A0304   &\hfil 5463  &\hfil $5505\pm34$&\hfil $75.9\pm0.5$&\hfil $0.190\pm0.002$&\hfil$2\pm2$ &\hfil 0.113 - 0.134  \\ 
          \hfil A0346 &\hfil 5877  &\hfil $5837\pm28$ &\hfil $88.3\pm1.3$&\hfil $0.137\pm0.001$&\hfil $69\pm5$&\hfil0.122 - 0.147 \\ 
          \hfil N6978 &\hfil 6115 &\hfil$5940\pm18$ &\hfil $89.1\pm0.8$&\hfil $0.122\pm0.001$&\hfil $80\pm3$ &\hfil 0.113 - 0.135 \\ 
          \hfil A0727 (0.25)  & \hfil 5880 &\hfil$5876\pm40$&\hfil $79.9\pm1.4$&\hfil $0.181\pm0.004$&\hfil $33\pm10$&\hfil 0.099 - 0.116 \\
          \hfil A0727 (0.75)  & \hfil 5880 &\hfil$5890\pm40$&\hfil $80.3\pm1.8$&\hfil $0.183\pm0.004$&\hfil $25\pm10$&\hfil 0.099 - 0.116 \\
           \hfil A0727 (Mean)  & \hfil 5880 &\hfil$5883\pm40$&\hfil $80.1\pm1.6$&\hfil $0.182\pm0.004$&\hfil $29\pm10$&\hfil 0.099 - 0.116 \\
           \hfil A0748 &\hfil 6107 & \hfil $6085\pm27$&\hfil $76.6\pm2.0$&\hfil$0.178\pm0.002$&\hfil $25\pm5$ &\hfil 0.091 - 0.105 \\
        \hfil A0806  &\hfil 6767  &\hfil $6737\pm22$&\hfil $90.00^{+0.00}_{-0.5}$&\hfil $0.143\pm0.001$&\hfil $55\pm8$ &\hfil 0.063 - 0.069 \\ 
        \hfil IT Cnc  &\hfil 6127 &\hfil $5995\pm30$&\hfil $77.2\pm0.7$&\hfil $0.123\pm0.002$&\hfil $39\pm9$ &\hfil 0.083 - 0.094  \\
        \hfil PS Com (0.25)&\hfil 5061&\hfil$4939\pm31$&\hfil $86.4\pm2.9$&\hfil $0.242\pm0.007$&\hfil $38\pm10$&0.136 - 0.166\\
        \hfil PS Com (0.75)&\hfil 5061&\hfil$4951\pm37$&\hfil $88.6\pm2.1$&\hfil $0.239\pm0.007$&\hfil $32\pm9$&0.136 - 0.166\\
        \hfil PS Com (Mean)   &\hfil 5061  &\hfil $4945\pm34$&\hfil $87.5\pm2.5$&\hfil $0.240\pm0.007$&\hfil $35\pm10$ &\hfil 0.136 - 0.166\\ 
           \hline
           
    \end{tabular}
    
    \end{table*}

  \subsection{O'Connell effect and the light curve solution}
  The light curves of contact binaries not infrequently show the two out of eclipse maxima as unequally high. The observations were initially thought to be related tidal distortion \citep{1906MNRAS..66..123R} or Hydrogen absorption \citep{1950Urani..21...58M}. The first major study of the phenomenon was carried out by \citet{1951PRCO....2...85O}. There is no single broadly applicable theory to explain the asymmetry principally because the O'Connell effect is known to be highly variable. The asymmetry may shift between the two maxima \citep{2022RAA....22c5024X}, the asymmetry shifts between bands, that is the first maxima maybe brighter in one colour band but fainter in another colour during simultaneous observations \citep{2011AJ....142..117S}, the O'Connell effect maybe transient or even reverse over very short time frames \citep{2022RAA....22b5005W, 2016NewA...47....3Z}. Given the observed variations in the O'Connell effect many theories have been proposed to explain it including starspots \citep{2009SASS...28..107W}, asymmetric distribution of circumference binary matter \citep{2003ChJAA...3..142L}, Coriolis heating of the photosphere \citep{1990ApJ...355..271Z} and gas stream impacts leading to an accretion hot spot \citep{2004A&A...423..607G, 1995IBVS.4177....1G}.  
  
  Incorporation of starspots in the derivation of light curve solutions is almost ubiquitous in systems exhibiting the O'Connell effect. The popularity of spotted solutions is most likely the result of their support with the widely used Wilson-Devenney modelling code. A well-placed spot will lead to a better fitting modelled light curve however if no justification is provided for the existence of such a spot(s) than confidence in the solution is actually reduced even if the curve fit look better. \citet{2018maeb.book.....P} in his book strongly advocates the use of starspots only if there is definitive proof, such as high-resolution Doppler spectroscopy, for the existence of spots. Similarly, \citet {2014ApJS..213....9D} argue against starspots as the cause for the O'Connell effect. One can understand such concerns when one examines the major issues that exist with the uniqueness of spotted solution and the strong correlations between geometric parameters and atmospheric spot parameters. As shown by \citet{1993A&A...277..515M} photometric analysis cannot distinguish between bright and dark spots being incorporated onto each component or the connecting neck. They find photometric solutions can vary by as much as 50\% in the mass ratio, fillout factor can change by almost 100\%, temperature difference can vary by over 200$\rm K$ and inclination can also change substantially. Such variations are continually being reported, \citet{2022PASA...39...57H} reported essentially indistinguishable fits for four different spot parameters. The main issue of concern is the different geometric solution each case produced. Similarly, \citet{2016AJ....151...69S} report similar fits with different spot parameters with different geometric parameters. Unfortunately, the popularity of spotted solutions remains because a well fitted curve looks aesthetically more pleasing even though as pointed out by authors including \citet{1994IAPPP..54....1H, 1997A&A...321..811O, 1992PASA...10...33A} that an infinite number of solutions may exist with equally good fits as there are infinite combinations of number, sizes, locations, and shapes of starspots. In reality, there cannot exist any single spotted solution and each equally well fitted solution will have a different geometric solution \citep{1999NewA....4..365E}.  The current interest in orbital stability of contact binary systems and the theoretical framework underpinning it requires a stable estimation of geometric parameters and given the problem with the uniqueness of spotted solutions and the potentially large variations in fitted geometric parameters it is difficult to accept such solutions for further analysis. For this reason, we chose to model the light curves of A0727 and PS Com twice firstly with maximum light at the first maximum (phase 0.25) and secondly with the maximum light set at the second maximum (phase 0.75). We adopted the mean of the two solutions to derive our geometric and absolute parameters. 

  \section{Absolute Parameters and Orbital Stability}

 Parts of the geometric light curve solution such as the mass ratio can be used to derive the mass of the secondary component ($M_2$) and from the period and Kepler's 3rd law the separation ($A$) between the components can be determined. The light curve solution also provides fractional radii of the components in different orientations and the geometric mean of these ($r_1, r_2$) is used to estimate the absolute radii of the components as $(R_1, R_2)$ = $A\times(r_1, r_2)$. As noted by \citet{2023RAA....23k5001W}, small variations as low as 200$K$ can lead to a greater than 10\% change in the estimation of the mass of the primary and subsequent greater discrepancy in the estimation of instability parameters. Absolute parameter estimations based on the light curve solution geometric elements are not temperature dependent as clearly shown by \citet{2023SerAJ.207...21W}. Additionally, black body estimations assume a spherical configuration, components of contact binary systems are highly distorted by the Roche geometry such that the mean radius of both components are larger than their main sequence equivalents \citep{2022JApA...43...94W}. For these reasons we prefer the use of observational and geometric elements from light curve solutions as a method of approximating absolute parameters. The absolute parameters are summarised in Table 4.

 Although it is well known that orbital instability is likely to occur when the mass ratio is low, it has only recently been shown that there is no global minimum mass ratio at which instability will ensue, instead the instability mass ratio ($q_{inst}$) is unique for each system \citep{2021MNRAS.501..229W}. The same study also provides two simple quadratic relations as follows:
 \begin{equation}
\label{eq:qinst-f1}
    q_{inst}=0.1269M_{1}^2-0.4496M_{1}+0.4403 (f=1)
\end{equation} 
and
\begin{equation}
\label{eq:qinst-f0}
  q_{inst}=0.0772M_{1}^2-0.3003M_{1}+0.3237 (f=0).  
\end{equation}
from which we can determine the instability mass ratio range from marginal contact (f=0) to full overcontact (f=1) depending on the mass of the primary component. We calculate the instability mass ratio range for each system as recorded in Table 3. We note that for three systems, A0346, V1359 Cas and N6798, the modelled mass ratio is within the instability mass ratio range and as such the systems would be classified as potential merger candidates. The remainder of the systems have mass ratios greater than the maximum instability mass ratio and therefore would be considered stable.

\begin{table*}

   \centering
\caption{Absolute parameters of the systems studied}
   \begin{tabular}{|c|c|c|c|c|c|c|}
    \hline
        \hfil Name &\hfil$M_1/M_{\sun}$ &\hfil  $M_2/M_{\sun}$ &\hfil $R_1/R_{\sun}$&\hfil$R_2/R_{\sun}$ &\hfil $A/R_{\sun}$ &\hfil $M_{V1}$ \\
        \hline
        \hfil N3560 &\hfil $1.18\pm0.02$ &\hfil $0.14\pm0.02$  &\hfil $1.42\pm0.05$&\hfil $0.68\pm0.03$&\hfil $2.43\pm0.08$&\hfil $3.68\pm0.01$ \\
        
        \hfil V1359 Cas   &\hfil $0.89\pm0.01$ &\hfil $0.11\pm0.01$ &\hfil $1.23\pm0.3$&\hfil $0.50\pm0.02$&\hfil $2.11\pm0.02$&\hfil $5.54\pm0.02$\\ 
         \hfil A0304   &\hfil $0.96\pm0.01$  &\hfil $0.17\pm0.02$&\hfil $0.97\pm0.01$&\hfil $0.45\pm0.01$&\hfil$1.83\pm0.02$ &\hfil $5.16\pm0.01$  \\ 
          \hfil A0346 &\hfil $0.86\pm0.01$  &\hfil $0.12\pm0.01$ &\hfil $1.10\pm0.01$&\hfil $0.49\pm0.02$&\hfil $1.90\pm0.02$&\hfil$5.77\pm0.02$ \\ 
          \hfil N6978 &\hfil $0.91\pm0.01$ &\hfil$0.11\pm0.01$ &\hfil $1.27\pm0.03$&\hfil $0.54\pm0.02$&\hfil $2.16\pm0.02$ &\hfil $5.72\pm0.02$ \\ 
           \hfil A0727  & \hfil $1.01\pm0.02$ &\hfil$0.18\pm0.02$&\hfil $1.12\pm0.02$&\hfil $0.53\pm0.02$&\hfil $2.07\pm0.03$&\hfil $4.70\pm0.02$ \\
           \hfil A0748 &\hfil $1.06\pm0.01$ & \hfil $0.15\pm0.01$&\hfil $1.10\pm0.01$&\hfil$0.52\pm0.01$&\hfil $2.04\pm0.02$ &\hfil $4.51\pm0.01$ \\
        \hfil A0806  &\hfil $1.31\pm0.02$  &\hfil $0.16\pm0.02$&\hfil $1.22\pm0.02$&\hfil $0.54\pm0.02$&\hfil $2.13\pm0.02$ &\hfil $3.58\pm0.02$ \\ 
        \hfil IT Cnc  &\hfil $1.13\pm0.01$ &\hfil $0.14\pm0.01$&\hfil $1.34\pm0.02$&\hfil $0.54\pm0.02$&\hfil $2.32\pm0.02$ &\hfil $3.97\pm0.01$  \\
        \hfil PS Com   &\hfil $0.77\pm0.01$  &\hfil $0.19\pm0.01$&\hfil $0.87\pm0.01$&\hfil $0.47\pm0.01$&\hfil $1.67\pm0.02$ &\hfil $6.48\pm0.01$\\ 
           \hline
           
    \end{tabular}
    
    \end{table*}

\section{Chromospheric Activity}
Loss of angular momentum is thought to be a major factor leading to orbital instability of contact binaries and magnetic breaking is likely a significant contributor \citep{2021AJ....162...13L}. Binary systems with outer convective envelopes almost always show signatures of magnetic activity with observable features such as starspots, spectra showing chromospheric emissions, and excess overall emissions at shorter wavelengths. In this section we report, for the first time, filling of the infrared Calcium absorption lines centered on $\lambda$8542  and $\lambda$8662 in low mass ratio contact binaries. In addition, we look at excess ultraviolet emissions in three systems observed by the Galaxy Evolution Explorer (GALEX) mission.

\subsection{CaII infrared filling and chromospheric activity}
The infrared CaII triplet (CaII IRT) lines at $\lambda$8498, $\lambda$8542 and $\lambda$8663 are a distinct spectral feature of cool low mass stars \citep{2017A&A...605A.113M}. Analysis of CaII infrared lines (usually limited to the $\lambda$8542 and $\lambda$8662 lines) is somewhat easier than the CaII lines in the bluer part of the spectrum as the continuum is much better-defined making normalization simpler \citep{2007A&A...466.1089B}. The CaII IRT has been used as a diagnostic indicator of chromospheric activity for some time. Unlike the wings of the CaII IRT which are quite extended and affected by multiple photospheric layers and the atmospheric temperature distribution \citep{2005A&A...430..669A}; the cores of the CaII IRT lines are due to the uppermost atmospheric layers (chromosphere), their central depression has been shown to be sensitive to the degree of activity and correlate well with the Mount-Wilson $R^{'}_{HK}$ activity indicator \citep{2000A&A...353..666C}.

Few authors have pointed to some problems with the use of CaII IRT lines for chromospheric diagnostics. The main concerns relate to significant dependence of the central depth due to rotation and to a lesser extent potential photospheric contributions \citep{2007A&A...466.1089B}. Both concerns can be addressed by using the subtraction technique whereby the spectrum of a carefully selected (matched for temperature and metallicity) comparison star is artificially broadened to the target star rotational velocity and then subtracted from the spectrum of the target star \citep{2017A&A...605A.113M}. The result is a purely activity related excess flux without photospheric or basal flux contributions. As noted above the wings of the CAII IRT lines are more dependent on non chromospheric atmospheric layers and as such only the central depth of the line cores is considered a chromospheric activity indicator. The central depth is usually reported as the normalised amplitude relative to the continuum \citep{2005A&A...430..669A}. 

The Large Sky Area Multi-Object Fibre Spectroscopic Telescope (LAMOST) has been conducting spectroscopic sky survey for over 10 years with regular data releases since 2013. All systems reported in this study, except V1359 Cas, were observed by the LAMOST survey and medium resolution spectra are available through data release 7 (DR7).  \citet{2023ApJS..265...61J} recently published a spectral library based on LAMOST spectra. As part of this library, they include a second library of over 1100 spectral templates which essentially represent smoothed averaged spectra of similar parameters. They suggest the secondary atlas can be used as a standard atlas for comparing LAMOST spectra. We artificially broaden the spectrum of a standard star from the atlas with similar parameters to the primary component of each system. We calculate the excess flux of the central depression as a "filling" of the normalised central flux defined as normalised flux amplitude (standard star) - normalised flux amplitude (primary component). A value greater than zero would indicate excess flux in the contact binary. Detailed procedure is described below.
\subsection{CaII IRT Excess Flux}
\subsubsection{Primary Component Spectrum and Broadening of the Standard Spectrum}
As noted above for all systems, except V1359 Cas, LAMOST spectra are available. Where more than one spectrum was available, we chose the one with the highest signal to noise ratio. 
The resolution of LAMOST spectra is relatively low and our study is limited to the $\lambda$8542 and $\lambda$8662 lines. The observed spectra is usually normalised to the average of the continuum a few angstrom on either side of the wings of the main absorption line \citep{2022AJ....164..202L}. In our case we normalise to a 5\AA \ region from 8525 - 8529\AA \ and 8561 - 8565\AA \ for the $\lambda$8542 line and because the $\lambda$8662 line demonstrates slightly asymmetric wings we used 8650 - 8654\AA  \ and 8676 - 8680\AA \ regions. As noted, the resolution of the LAMOST spectra is low, rather than using the mean of the flux to calculate the continuum which could be influenced significantly by a single outlier reading over the short continuum length, we first "smoothed" the wings and continuum of the LAMOST spectrum while maintaining the observed absorption depth by applying a locally estimated non-parametric regression method (LOESS) \citep{7df9e592-ea23-368d-845f-c8feba60f603}. The spectra were then normalised using the WinMK24a package \citep{2003gafe.conf..125M}. 

As noted above central depression filling of the infrared line core can be achieved through subtraction of a matched standard spectrum that has been corrected for broadening. The LAMOST survey not only provides an estimate of the effective temperature but also the metallicity for each system. We use the LAMOST temperature and metallicity to select the nearest match from the standard sample atlas (grouped by temperature and metallicity) as our standard star. In the case of A0346 the LAMOST survey does have a spectrum of the system but does not provide an effective temperature or metallicity. In this case we use the GAIA EDR3 estimate of for the temperature and metallicity. The GAIA metallicity was corrected as described by \citet{2023A&A...674A..27A}. Contact binary orbits are thought to be synchronous and as such it is possible to calculate the rotation speed of the primary from the period and estimated radius. Our estimates of the absolute radii are summarised in Table 4. The spectrum of the standard star was then broadened by applying the estimated rotational velocity and inclination with the recently published script \citep{2023RNAAS...7...91C}. 

\subsubsection{Central Depression Filling}
The excess flux of the CaII IRT absorption lines was calculated as the difference between the normalised amplitude of the central depression of the broadened standard star and the normalised amplitude of the central depression of the contact binary as advocated by \citet{2007A&A...466.1089B}.  Given slight variations due to radial velocity and red shift we shifted the template spectra to keep the corresponding spectral line wavelengths consistent, as per \citep{2023MNRAS.519.5760L}. A positive result for excess filling was confirmed when the central depression of the broadened lines of the standard star was deeper than that of the lines of the primary component. Calculations were carried out for the $\lambda$8542 and $\lambda$8662 lines in all cases except V1359 Cas where no LAMOST spectrum was available. Effects of the broadening on the filling of the central depression of the standard star and excess filling due to chromospheric activity are illustrated in Figure 2. Table 5 summarises the temperature and metallicity characteristics of the contact binary systems and the selected standard stars along with central depression filling for $\lambda$8542 and $\lambda$8662 lines. We note that the reported metallicity for A0806 at -1.62 is very low with estimated temperature 6767$K$. We could not find a standard star with such low metallicity and similar temperature. We chose the closest match with metallicity -0.55 and temperature 6751$K$.

 \begin{figure}
    \label{fig:F2}
	\includegraphics[width=\columnwidth, height = 22cm]{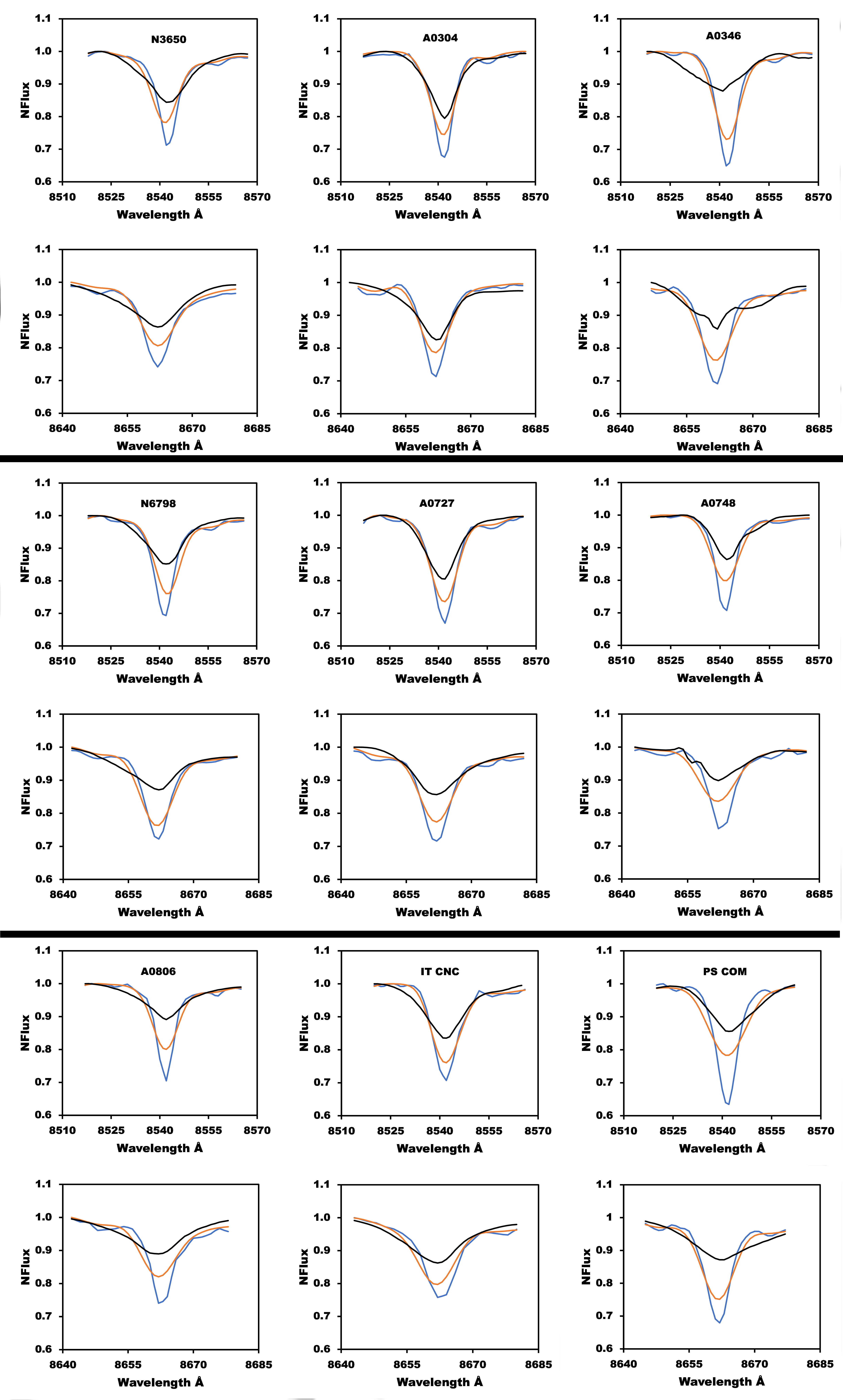}
    \caption{Broadening and filling of infrared calcium lines. The blue line is the un-broadened spectrum of the standard star. The orange the broadened line (standard star) and the black line is the spectral line of the contact binary. The top panel illustrates the $\lambda 8542$ line and the bottom panel the $\lambda 8662$ line. The vertical axis labelled NFlux represents the normalised flux.}
    \end{figure}

As noted above observational signals of excess chromospheric activity are common in contact binary systems. The CaII IRT lines have been shown to be very useful in detection of increased chromospheric activity, so it is not surprising that all our systems demonstrate some degree of filling of the central depression in $\lambda$8542 and $\lambda$8662 lines. The extent of filling for each line was similar with the mean ($\pm SD$) normalised flux filling of $0.091\pm0.036$ for the $\lambda$8542 line and $0.082\pm0.029$ for the $\lambda$8662 line. The usual method of equivalent width assessment for determining excess flux is not suited for the CaII IRT lines as the broadened wings of the lines are highly dependent on denser photospheric elements and difficult to separate from less dense upper atmospheric (chromospheric) emissions. The line cores and in particular the central depressions are less dependent on photospheric emissions \citep{2005A&A...430..669A}. We are not suggesting that only low mass ratio contact binaries have excess CaII IR filling. It is possible, in fact likely, that many contact binaries will demonstrate this feature. The finding is another marker of increased chromospheric activity that can be looked for in the absence other markers or as supporting evidence for enhanced chromospheric activity.
\begin{table*}
\centering
 \caption{Temperature and metallicity of the systems observed from LAMOST DR 7 with temprature and metallicity of the selected standard star. (CN) = catalogue number from the atlas of standard star. Last two columns summarise the normalised flux filling for each system at $\lambda 8542$ and $\lambda 8662$.\\A0346$^*$ - No LAMOST parameters available - GAIA parameters used}
\begin{tabular}{|c|c|c|c|c|}
    \hline
        \hfil Name &\hfil System &\hfil  Standard Star &\hfil$ \lambda 8542$&\hfil $\lambda 8662$ \\
        \hline
        \hfil &\hfil Temp(K) / [Fe/H] &\hfil Temp(K) / [Fe/H] (CN) &\hfil Filling&\hfil Filling \\
        \hline
        \hfil N3560 &\hfil 6568 / -0.213 &\hfil  6547 / -0.171 (526)  &\hfil 0.063&\hfil 0.053 \\
        
        \hfil A0304   &\hfil 5463 / -0.665  &\hfil 5454 / -0.547 (629)&\hfil0.054&\hfil 0.057 \\ 
          \hfil A0346$^*$ &\hfil 5877 / -0.113  &\hfil 5865 / -0.151 (583) &\hfil 0.161&\hfil 0.095 \\ 
          \hfil N6978 &\hfil 6115 / -0.228 &\hfil 6109 / -0.256 (556) &\hfil 0.090 &\hfil 0.121 \\ 
           \hfil A0727  & \hfil 5880 / -0.06 &\hfil 5911 / -0.028 (292)&\hfil 0.068&\hfil 0.078 \\
           \hfil A0748 &\hfil 6107 / -0.646 & \hfil 6157 / -0.631 (188)&\hfil 0.061&\hfil0.059 \\
        \hfil A0806  &\hfil 6767 / -1.62  &\hfil 6751 / -0.550 (505)&\hfil0.096&\hfil0.072 \\ 
        \hfil IT Cnc  &\hfil 6127 / -0.016 &\hfil 6115 / -0.079 (469)&\hfil 0.076&\hfil 0.067  \\
        \hfil PS Com   &\hfil 5061 / 0.054  &\hfil 5066 / 0.012 (713)&\hfil 0.146&\hfil 0.140\\ 
           \hline
           
    \end{tabular}
   
    \end{table*}

\subsection{Ultraviolet Emissions}
Another hallmark of increased chromospheric activity is increased ultraviolet (UV) emissions. As with assessment of the CaII IRT filling, the UV emissions can be contaminated by photospheric emissions and only emissions from the far UV part of the spectrum (below 1800\AA) can be regarded as being mainly from chromospheric activity \citep{2010PASP..122.1303S}. Like correlation between $R^{'}_{HK}$ and central depression of CaII IRT lines, \citet{2010PASP..122.1303S} explored the relationship between $R^{'}_{HK}$ and the far ultraviolet (FUV) magnitudes from the Galaxy Evolution Explorer (GALEX) mission. They defined a UV colour excess $\Delta(m_{FUV}-B)$ as follows:

\begin{equation}
    (m_{FUV}-B)_{base} = 6.73(B-V) + 7.43,
\end{equation}
which defines $m_{FUV}-B$ for stars with the weakest emissions and low activity;
and
\begin{equation}
    \Delta(m_{FUV}-B) = (m_{FUV}-B) - (m_{FUV}-B)_{base}
\end{equation}
where $m_{FUV}$ represents the GALEX far ultraviolet magnitude.

They correlated the UV colour excess with $R^{'}_{HK}$ and found that active stars have a colour excess below -0.5 (usually below -1.0) while inactive stars have colour excess well above -0.5.

Three of the systems presented in the current study, A0748, A0806 and N3650 were observed by the GALEX mission with measured $m_{FUV}$. We use the published values for $B-V$ for the systems to calculate the UV colour excess for the three system as follows -2.31, -3.61 and -2.16 respectively. All three have a UV colour excess well below -0.5 again pointing to significant chromospheric activity.

\section{Summary and Concluding Remarks}

Orbital stability of contact binaries has received significant attention particularly since the confirmation that Nova Sco 2008 (=V1309 Sco), a red nova, was in fact a merger event of a low mass ratio contact binary \citep{2011A&A...528A.114T}. It was recognised some time ago that merger events are only likely if the mass ratio of a contact binary system was low. For quite some time investigators attempted to define a global minimum mass ratio at which merger was likely. Unfortunately, systems with mass ratios smaller than the proposed global minimum were being continually identified placing the concept of a global mass ratio into the doubtful column. Recently \citet{2021MNRAS.501..229W} published revised orbital stability criteria and found that there is not one global minimum mass ratio at which merger would occur, instead each system has its unique potential merger mass ratio dependent on the mass of the primary component. They showed that for systems where the primary is of $0.6M_{\sun}$ the instability mass ratio is more than 0.2 while for a system where the primary is of $1.6M_{\sun}$ the instability mass ratio is closer to 0.05.

We selected ten contact binary systems from various sky surveys whose photometry suggested that they were likely to have low mass ratios and as such potentially merger candidates. We undertook ground-based observations and analysis of the light curves and confirm that all ten are indeed of low mass ratio ranging from 0.122 - 0.24. Using the instability criteria from \citet{2021MNRAS.501..229W} we find that three of our systems A0346, V1359 Cas and N6798 are potentially unstable with mass ratios in the instability range. 

Loss of angular momentum due to magnetic breaking is considered as a potential mechanism leading to eventual orbital instability. Increased magnetic activity can be suspected in the presence of increased chromospheric activity. Although many indicators of increased chromospheric have been used, spectroscopic excess of certain emission and absorption lines and excess emission at higher energy (ultraviolet) have been among the most widely used. Nine of the ten systems reported in this study had available medium resolution spectra from the LAMOST survey. We used these along with matched standard stars to measure the excess flux for two of the three CaII IRT lines. We chose to look at the core of the absorption lines rather than the full width as the line wings are more influenced by photospheric contamination. The amplitude of the core of CaII IRT lines has been shown to be a good marker of increased chromospheric activity and all nine of the systems reported in this study with available spectra have significant filling of the central depression compared to matched standard stars. The extent of central depression filling was similar for the two lines examined. In addition to infrared spectroscopic evidence for increased chromospheric activity we were able to look at increased ultraviolet emissions for three systems observed with GALEX mission. All three systems show a UV colour excess well below the level indicative of a chromospherically active star and when combined with increased CaII IRT flux strengthen the case for increased chromospheric activity in contact binary stars. 

The main light curve feature attributed to possible increased chromospheric activity has been the O'Connell effect and that only if one attributes the observed asymmetry in the maxima to starspots. There exists ample evidence to indicate that light curve solutions incorporating starspots may lead to erroneous estimations of geometric and absolute parameters. Lack of confidence in the light curve solution would then flow onto reduced confidence in orbital stability and spectroscopic broadening considerations. Due to issues with the uniqueness of the light curve solution incorporating starspots, we adopted solutions without spots for two systems, A0727 and PS Com, which show significant O'Connell effect. A0727 demonstrates central depression filling somewhat less than the mean of the nine systems presented while PS Com demonstrates greater filling than the mean filling. From the small sample presented here all low mass ratio contact binary appear to have increased chromospheric activity without necessarily having significant light curve features. We agree with authors such as \citet{2018maeb.book.....P} that starspots should only be considered as part of the light curve solution if there is significant other evidence such as Doppler spectroscopy confirming their existence on a particular star. Otherwise, the light curve fit may well look better however with high risk of an erroneous overall solution.

Two of the systems presented in this study, N6798 and N3650 have previously published light curve solutions. \citet{2022AJ....164..202L} analysed multi-band and TESS photometry for both systems. For N3650 they report a mass ratio of 0.142 with fillout 96.3\% and inclination of 77.8$^{\circ}$ from ground-based photometry and 0.146 with fillout 99.7\% and inclination 77.2$^{\circ}$ from TESS photometry. They report no O'Connell effect. For N6798 they analysed ground-based and TESS photometry from two sectors. They observed the O'Connell effect in the ground-based observations and in one of the TESS sectors. They report a mass ratio of 0.128 with a fillout of 57.3\% and inclination of 87.5$^{\circ}$ for a spotted solution from ground based observations and mass ratios of  0.136, fillout 97.3\% and inclination of 85.9$^{\circ}$ for the TESS sector with no O'Connell effect and mass ratio of 0.135, fillout 84.6\% and inclination 87.6$^{\circ}$ for a spotted solution of the TESS photometry with detectable O'Connell effect. These results again demonstrate potential pitfalls of introducing spots with respect to the geometric light curve solutions. For N3650 there is no significant difference with respect to the three main geometric parameters between the ground based and TESS photometry. In the case of N6798 there is over a 6\% difference in the mass ratio, 70\% difference in the fillout and about a 2\% difference in inclination between the spotted and unspotted solutions. The difference in the estimated mass ratio is particularly concerning as it has a very direct effect on the determination of the absolute parameters. Our solution for N3650 with a mass ratio of 0.147, fillout 92\% and inclination 79.3$^{\circ}$ is in general agreement. Our solution for N6798 with no observed O'Connell effect finds the mass ratio lower at 0.122 with a fillout of 80\% and high inclination at 89.1$^{\circ}$. Clearly the system has some unusual features and ongoing observations are warranted.

\section*{Acknowledgements}
 
\noindent This research has made use of the SIMBAD database, operated at CDS, Strasbourg, France.\

\noindent This work makes use of observations from the Las Cumbres Observatory global telescope network\

\noindent Authors acknowledge the assistance of Joel Balzan and Evan Crawford from Western Sydney University in implementation of the broadening script.\

\noindent  OV and PK acknowledge support by the Astronomical station Vidojevica, funding from the Ministry of Science, Technological Development and Innovation of the Republic of Serbia (contract No. 451-03-66/2024-03/200002), by the EC through project BELISSIMA (call FP7-REGPOT-2010-5, No. 265772).\

\noindent AP was financed by Silesian University of Technology Statutory Activities Grant No. BK-250/RAu-11/2024\\

\bibliographystyle{raa}
\bibliography{bibtex}

\end{document}